\documentclass[slac_one]{revtex4}
\usepackage{graphicx}
\usepackage{fancyhdr}
\renewcommand{\headrulewidth}{0pt}
\renewcommand{\footrulewidth}{0pt}

\newcommand{\Ds}{D_s^+}
\newcommand{\Dp}{D^+}
\newcommand{\fd}{f_{\Dp}}
\newcommand{\fds}{f_{\Ds}}
\newcommand{\etal}{{\it et al.}}
\newcommand{\mbc}{m_{\rm BC}}

\begin{document}

\title{Measurements of $D$ and $D_s$ decay constants at CLEO} 

%

\author{L.~M.~Zhang}
\affiliation{Physics Department, Syracuse University, Syracuse, N.
Y. 13244, USA}

\begin{abstract}
Using CLEO data collected at 3370 MeV and 4170 MeV, we determine
$f_{\Dp} = (205.8\pm8.5\pm2.5)$ MeV and an interim preliminary value
of $f_{\Ds} = (267.9\pm8.2\pm3.9)$ MeV, where both results are
radiatively corrected. They agree with the recent most precise
unquenched Lattice-QCD calculation for the $\Dp$, but do not for the
$\Ds$. Several consequences are discussed.
\end{abstract}

\maketitle

\thispagestyle{fancy}


\section{Introduction} 
\begin{figure}[!htbp]
\includegraphics[width=0.4\textwidth]{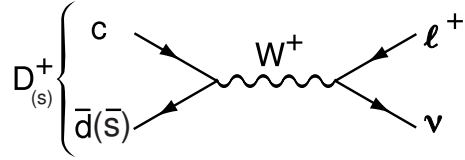}
\label{diagram}\caption{Decay diagram for $D_{(s)}^+\to \ell^+ \nu$.
}
\end{figure}

Leptonic decay $D_{(s)}^+\to \ell^+ \nu$ is described by the
annihilation of the initial quark-antiquark pair into a virtual
$W^+$ that materializes as a pair (Fig. \ref{diagram}).

The decay rate is give by
\begin{equation}\label{eq:equ_rate}
\Gamma(D_{(s)}^+\to \ell^+ \nu) =
\frac{G_F^2}{8\pi}f_{D_{(s)}^+}m_\ell^2m_{D_{(s)}^+}\left(1-\frac{m_\ell^2}{m_{D_{(s)}^+}^2}\right)|V_{cd(s)}|^2,
\end{equation}
where $f_{D_{(s)}^+}$ is decay constant \cite{Rosner-Stone}, related
to the overlap of the heavy and light quark wave-function at zero
spatial separation, $G_F$ is the Fermi constant, $m_{D^+_{(s)}}$ is
the $D_{(s)}^+$ mass, $m_\ell$ is the final state charged-lepton
mass, and $V_{cd(s)}$ is a CKM matrix element, taken as
$V_{cd}=V_{us}=0.2256$ and $V_{cs}=V_{ud}=0.9742$. Thus, measurement
of purely leptonic decays allow a determination of the decay
constant $f_{D_{(s)}^+}$.

Meson decay constants in the $B$ system are used to translate
measurements of $B\bar{B}$ mixing to CKM matrix elements. Currently,
it is not possible to determine $f_{B}$ accurately from leptonic $B$
decays, so theoretical calculations of $f_{B}$ are used. Since the
$B_s^0$ meson does not have $\ell^+\nu$ decays, it will never be
possible to determine $f_{B_s}$ experimentally, thus theory must be
relied on. If calculations disagree on $D$ mesons, they may be
questionable on $B$ mesons. If, on the other hand new physics is
present, we need understand how it effects SM based predictions of
the $B$ decay constants. Decay constants can be calculated using
lattice quantum-chromodynamics (LQCD). Recent calculation from
Follana \etal~using an unquenched LQCD predicts $\fd = (207\pm4)$
MeV and $\fds=(241 \pm 3)$ MeV \cite{Lat:Follana}.

We use the reactions $e^+e^- \to D^-D^+$, and $e^+e^-\to
D_s^{*-}D_s^+$ or $D_s^{-}D_s^{*+}$. The $D^+$ is studied at 3770
MeV using 818 pb$^{-1}$ of data \cite{prd-fd}. And the $\Ds$ is
studied at 4170 MeV, using 424 pb$^{-1}$ for $\mu^+ \nu$ and $\tau^+
\nu$, $\tau^+ \to \pi^+\bar{\nu}$, and 300 pb$^{-1}$ for $\tau^+
\nu$, $\tau^+ \to e^+ \nu \bar{\nu}$. (Eventually CLEO will present
results using 600 pb$^{-1}$.)

\section{\boldmath{$D^+\to \ell^+ \nu$}}
\begin{figure}[h]
\centering
\includegraphics[width=0.45\textwidth]{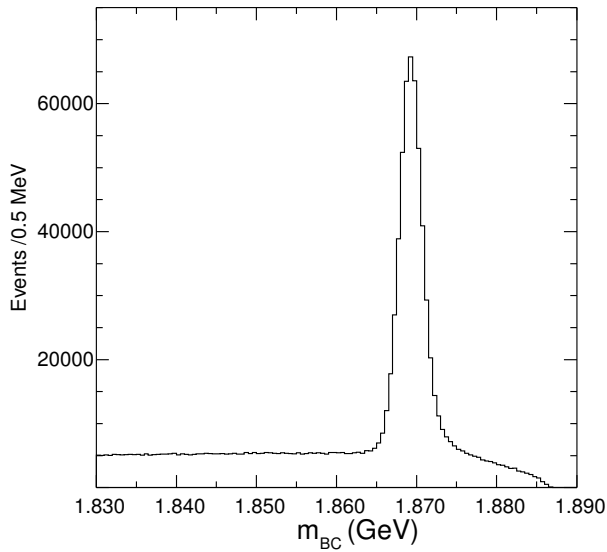}
\caption{The beam-constrained mass distributions summed over $D^-$
decay candidates in the final states: $K^+ \pi^- \pi^-$, $K^+ \pi^-
\pi^- \pi^0$, $K_S\pi^-$, $K_S \pi^-\pi^-\pi^+$, $K_S\pi^- \pi^0$
and $K^+K^-\pi^-$.} \label{mbc_all}
\end{figure}

We fully reconstruct $D^-$ \cite{cc} as a tag and examine the
properties of the other $\Dp$, which can be found even if there is a
missing neutrino in the final state. This method is called as
``double tag" technique. To reconstruct $D^-$ tags we require that
the tag candidates have a measured energy consistent with the beam
energy, and have a ``beam constrained mass", $\mbc$, consistent with
the $D^-$ nominal mass \cite{PDG}, where $\mbc=\sqrt{E_{\rm
beam}^2-\left(\sum_i\mathbf{p}_i\right)^2}$, $E_{\rm beam}$ is the
beam energy and $i$ runs over all the final state particles
three-momenta. Fig.~\ref{mbc_all} shows the $m_{BC}$ distribution
summed over all the decay modes we use for tagging. Selecting events
in the mass peak we count 460,055$\pm$787 signal events over a
background of 89,472 events.

We then search for signal events with one and only one additional
track with opposite sign of charge to the tag, not identified as
kaon. The track must make an angle $>25.8^\circ$ with respect to the
beam line (90\% of the solid angle), and in addition we require that
there not be any photon detected in the calorimeter with energy
greater than 250 MeV. The latter selection can highly suppress $\Dp
\to \pi^+ \pi^0$ background. We separate these events into two
cases, where case (i) refers to muon candidates depositing $<300$
MeV, characteristic of 98.8\% of muons and case (ii) is for
candidates depositing $>300$ MeV, characteristic of 45\% of the
pions, those that happen to interact in the calorimeter and deposit
significant energy.

We look for $D^+\to\mu^+\nu$ by computing the square of the missing
mass
\begin{equation}
{\rm MM}^2=\left(E_{\rm beam}-E_{\mu^+}\right)^2-\left(-{\bf
p}_{D^-} -{\bf p}_{\mu^+}\right)^2, \label{eq:MMsq}
\end{equation}
where ${\bf p}_{D^-}$ is the three-momentum of the fully
reconstructed $D^-$, and $E_{\mu^+}({\bf p}_{\mu^+})$ is the energy
(momentum) of the $\mu^+$ candidate. The signal peaks at zero for
$\mu^+\nu$ and is smeared toward more positive values for
$\tau^+\nu$, $\tau^+\to\pi^+\overline{\nu}$.

\begin{figure}[!hbtp]
\centering
\includegraphics[width=0.5\textwidth,height=0.37\textheight]{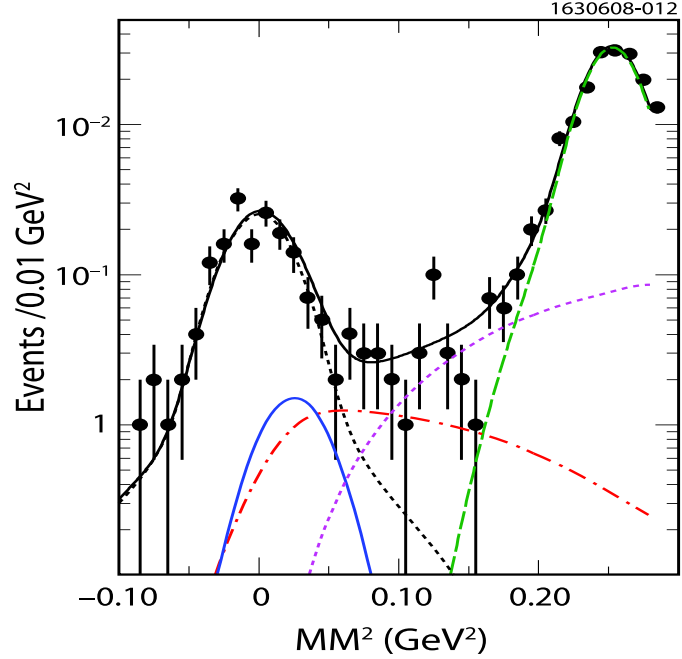}
\caption{Fit to the MM$^2$ for case (i). The points with error bars
 show the data. The black (dashed) curve centered at zero
 shows the signal $\mu^+\nu$ events. The dot-dashed (red) curve that peaks around
 0.05 GeV$^2$ shows the $D^+\to\tau^+\nu$, $\tau^+\to\pi^+\bar{\nu}$
 component.  The solid (blue) Gaussian shaped
 curve centered on the pion-mass squared shows the residual
 $\pi^+\pi^0$ component. The dashed (purple) curve that falls to
 zero around 0.03 GeV$^2$ is the
 sum of all the other background components, except the $\overline{K}^0\pi^+$
 tail which is shown by the long-dashed (green) curve that peaks up at
 0.25 GeV$^2$. The solid (black) curve is the sum of all the
 components.} \label{case1-taunufix}
\end{figure}

The fit to the case (i) MM$^2$ distribution shown in
Fig.~\ref{case1-taunufix} contains separate shapes for signal,
$\pi^+\pi^0$, $\overline{K}^0\pi^+$, $\tau^+\nu$ ($\tau^+\to
\pi^+\bar{\nu})$, and a background shape describing three-body
decays. Here we assume the SM ratio of 2.65 for the ratio of the
$\tau^+\nu/\mu^+\nu$ component and constrain the area ratio of these
components to the product of 2.65 with ${\cal{B}}(\tau^+\to
\pi^+\bar{\nu}$)=(10.90$\pm$0.07)\%~\cite{PDG} and the 55\%
probability that the pion deposits $<$300 MeV in the calorimeter.
The $\pi^+\pi^0$ background are fixed at 9.2 events obtained from
Monte Carlo (MC) simulation. The normalizations of the signal,
$\overline{K}^0\pi^+$, and 3-body background are allowed to float.
The $\overline{K}^0\pi^+$ shape is obtained from double tag events
of $D^0\to K^-\pi^+, \overline{D}{}^{0} \to K^+\pi^-$ where we
ignore one kaon to calculate the MM$^2$. All other shapes are
obtained from the MC simulation.

The fit yields 149.7$\pm$12.0 $\mu^+\nu$ signal events and 25.8
$\tau^+\nu$, $\tau^+\to\pi^+\bar{\nu}$ events (for the entire MM$^2$
range). We also perform the fit allowing the $\tau^+\nu$,
$\tau^+\to\pi^+\bar{\nu}$ component to float.  Then we find
153.9$\pm$13.5 $\mu^+\nu$ events and 13.5$\pm$15.3 $\tau^+\nu$,
$\tau^+\to\pi^+\bar{\nu}$ events, compared with the 25.8 we expect
in the SM. Performing the fit in this manner gives a result that is
independent of the SM expectation of the $D^+\to\tau^+\nu$ rate. To
extract a branching fraction, in either case, we subtract off
2.4$\pm$1.0  events found from simulations and other studies to be
additional backgrounds, not taken into account by the fit.

We find ${\cal{B}}(D^+\to\mu^+\nu)=(3.82\pm 0.32\pm 0.09)\times
10^{-4}.$ The decay constant $f_{D^+}$ is then obtained from
Eq.~(\ref{eq:equ_rate}) using 1040$\pm$7 fs as the $D^+$ lifetime
\cite{PDG} and 0.2256 as $|V_{cd}|$. Our final result, including
radiative corrections is
\begin{equation}
f_{D^+}=(205.8\pm 8.5\pm 2.5)~{\rm MeV}~.
\end{equation}

\section{\boldmath{$\Ds \to\ell^+ \nu$}}
\begin{figure}[!htbp]
\label{2d}
\includegraphics[width=0.4\textwidth]{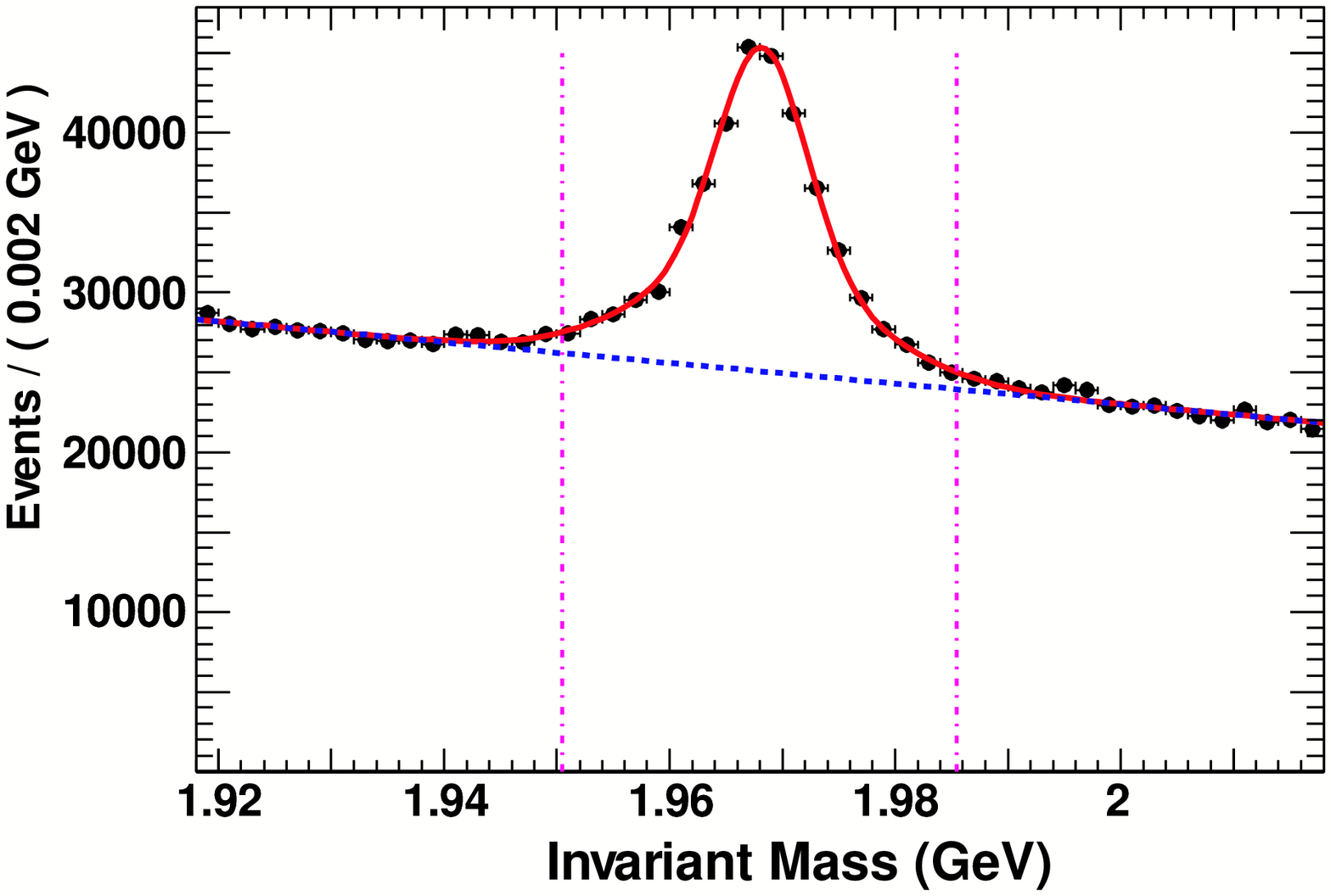}%
\includegraphics[width=0.4\textwidth]{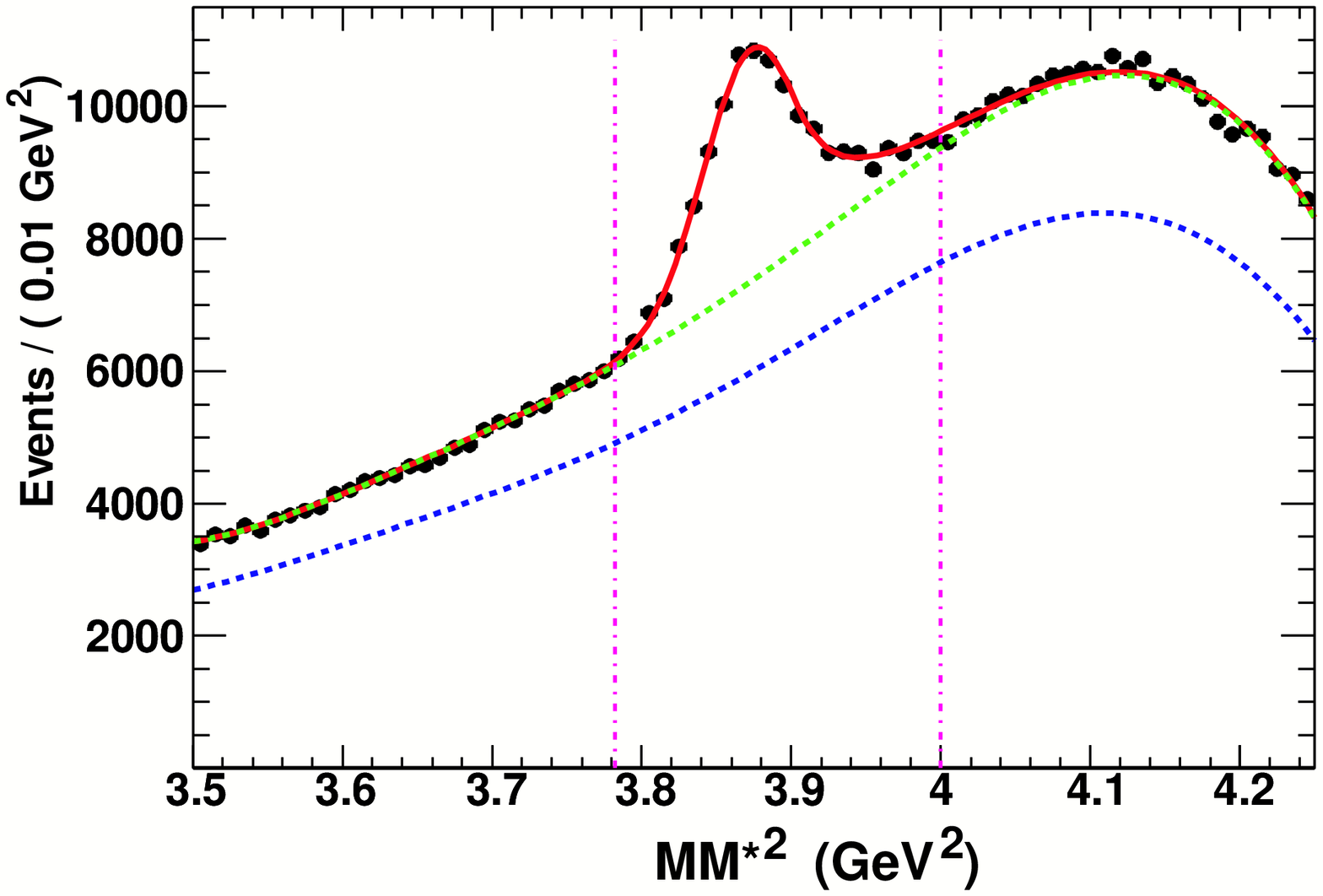}
\caption{Distributions of the invariant mass (left) in MM$^{*2} \in
[3.5, 4.25]$ GeV$^2$ and MM$^{*2}$ (right) in the invariant mass
$|M_{D_s}-1968.3$ MeV$|<17.5$ MeV in the final states: $K^+K^-\pi^-
$, $K_S K^-$, $\eta\pi^-$; $\eta\to\gamma\gamma$, $\eta'\pi^-$;
$\eta'\to\pi^+\pi^-\eta$, $\eta\to\gamma\gamma$, $\phi\rho^-$;
$\phi\to K^+K^-$, $\rho^-\to \pi^-\pi^0$, $\pi^+\pi^-\pi^-$,
$K^{*-}K^{*0}$; $K^{*-}\to K_S^0\pi^-$, ${K}^{*0}\to K^+\pi^-$,
$\eta\rho^-$; $\eta\to\gamma\gamma$, $\rho^-\to \pi^-\pi^0$, and
$\eta'\pi^-$; $\eta'\to\pi^+\pi^-\gamma$. Data (point) and fit
function (red curve) are shown. The dashed (blue) curve corresponds
to the fake $D_s$ background, and the dashed (green) curve to the
random photon background.}
\end{figure}

To reconstruct the tag, the difference here than the $D^+$ case is
that we need to include an additional photon from $D_s^{*}$ in the
tag. The $D_s^-$ tags we reconstructed can either from directly
produced $D_s^-$ mesons or those that result from the decay of
$D_s^{*}$ mesons. We calculate the missing mass squared MM$^{*2}$
recoiling against the photon and the $D_s^-$ tag,
\begin{equation}
{\rm MM}^{*2} = (E_{\rm CM}-E_{D_s}-E_{\gamma})^2-(\mathbf{p}_{\rm
CM}-\mathbf{p}_{D_s}-\mathbf{p}_{\gamma})^2,
\end{equation}
here $E_{\rm CM}$ (${\bf p}_{\rm CM}$) is the center-of-mass energy
(momentum), $E_{D_s}$ (${\bf p}_{D_s}$) is the energy (momentum) of
the fully reconstructed $D_s^-$ tag, and $E_{\gamma}$ (${\bf
p}_{\gamma}$) is the energy (momentum) of the additional photon. We
determine number of tags by simultaneously fit to the invariant mass
($M_{D_s}$) and MM$^{*2}$, shown in Fig. \ref{2d}. The signal is fit
to a sum of two Gaussians for the $M_{D_s}$ and a Crystal Ball
function for the MM$^{*2}$. The tail parameters of the Crystal Ball
function are obtained from fully reconstructed $D_s^*D_s$ events.
The background has two components: either comes from the background
under the invariant mass peak (fake $D_s$), or is due to random
photon combinations. The former background is modeled by linear
function for the mass and 5th order Chebyshev polynomial for the
MM$^{*2}$. Since the latter background is from a true $D_s$, its
mass distribution has the same shape as that from the signal, while
its MM$^{*2}$ is modeled by another 5th order Chebyshev polynomial
function. The total number of single tags is $30848\pm695\pm925$ in
the invariant mass signal region ($\pm17.5$ MeV from the nominal
$D_s$ mass) and MM$^{*2}\in$[3.782, 4.0] GeV$^2$.

\begin{figure}[h]
\centering
\includegraphics[width=0.45\textwidth]{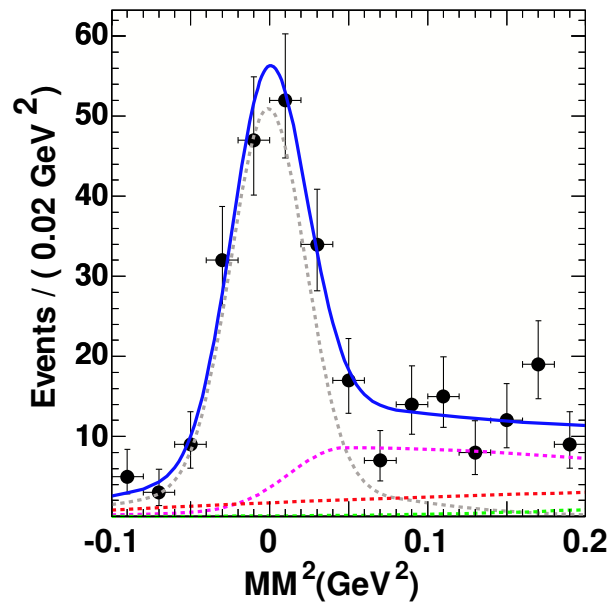}
\caption{The MM$^2$ distribution. The dashed (grey) Gaussian shaped
curve peaked near zero is the $\mu^+\nu$ component, while the dashed
(purple) curve that rises sharply from zero and then flattens out
shows the $\tau^+\nu$ component. The two lines are background
components. The solid curve shows the sum. }
\label{mms-fitall_38-47}
\end{figure}

Similarly as the $D^+$, we reconstruct the signal side. We veto
events with an extra neutral energy cluster $>300$ MeV (it is 250
MeV in the $D^+$ case). It is highly effective in reducing
backgrounds, especially for $\Ds \to \pi^+\pi^0$, $\eta\pi^+$ and
the processes $D^{(*)}\overline{D}{}^{(*)}$. The missing mass
squared, MM$^2$, evaluated by taking into account the observed
$\mu^+$, $D_s^-$, and photon should peak at zero;
\begin{equation}
{\rm MM}^{2} = (E_{\rm
CM}-E_{D_s}-E_{\gamma}-E_{\mu})^2-(\mathbf{p}_{\rm
CM}-\mathbf{p}_{D_s}-\mathbf{p}_{\gamma}-\mathbf{p}_{\mu})^2.
\end{equation}
We also make use of a set of kinematical constraints and fit each
event to two hypotheses one of which is that the $D_s^-$ tag is the
daughter of a $D_s^{*-}$ and the other that the $D_s^{*+}$ decays
into $\gamma\Ds$, with the $\Ds$ subsequently decaying into
$\mu^+\nu$. In addition, we constrain the invariant mass of the
$D_s^-$ tag to the know $D_s$ mass. This gives us a total of 7
constraints. The missing neutrino four-vector needs to be
determined, so we are left with a three-constraint fit. We perform a
standard iterative fit minimizing $\chi^2$. We choose the fitted
MM$^2$ from the hypothesis giving the smaller $\chi^2$. The MM$^{2}$
distribution from data is show in Fig. \ref{mms-fitall_38-47}. After
fixing the ratio of $\tau^+\nu/\mu^+\nu$ to the SM value we find
$f_{\Ds}=(268.2\pm9.6\pm4.4)$ MeV.

We can also use the decay mode $\tau^+\to e^+\nu\overline{\nu}$.
This result has already been published. \cite{CLEOtaunu} The
technique here is to use only three tagging modes: $\phi\pi^-$,
$K^-K^{*0}$ and $K^0_SK^-$, to ensure that the tags are extremely
clean. Then events with an identified $e^+$ and no other charged
tracks are selected. Any energy not associated with the tag decay
products is tabulated. Those events with small extra energy below
400 MeV are mostly pure $D_s^+\to\tau^+\nu$ events. After correcting
for efficiencies and residual backgrounds we find $f_{D_s^+}=(273\pm
16 \pm 8)$ MeV.

\section{Conclusions}

The preliminary CLEO average is $f_{D_s^+}=(267.9\pm 8.2 \pm 3.9)$
MeV (radiatively corrected). Averaging in the Belle result
\cite{Belle-munu} $f_{D_s^+}=(269.6\pm 8.3)$ MeV, which differs from
the Follana \etal~calculation~\cite{Lat:Follana} by 3.2 standard
deviations, while the result for $f_{D^+}=(205.8\pm 8.5 \pm 2.5)$
MeV is in good agreement. This discrepancy could be due to physics
beyond the standard model \cite{Dobrescu-Kron}, or systematic
uncertainties that are not understood in the LQCD calculation or the
experimental measurements, or unlikely statistical fluctuations in
the experimental measurements or the LQCD calculation. Fits to the
CKM matrix parameters use theoretical predictions of
$f_{B_s}$/$_{B_d}$. As similar calculations are used for
$f_{B_s}$/$_{B_d}$, we need to be concerned with them.

\section*{Acknowledgments}
I thank the U. S. National Science Foundation for support. Excellent
conversations were had with P. Chang, A. Kronfeld and S. Stone.


\begin{thebibliography}{9}
\bibitem{cc}
For notational convenance, the $D^-$ is referred to for the fully
reconstructed tag, although $D^+$ states are also used.
\end{thebibliography}
\end{document}